\documentclass[twocolumn,showpacs,amsmath,amssymb]{revtex4}

\usepackage{graphicx}
\usepackage{multirow}
\usepackage{epstopdf}
\usepackage{array}
\usepackage{longtable}
\usepackage{rotating,booktabs}
\usepackage{booktabs,threeparttable}
\usepackage{bm}
\usepackage{CJK}
\usepackage[marginal]{footmisc}
\usepackage{amsmath}
\usepackage{float}
\makeatletter

\newcommand{\Rmnum}[1]{\expandafter\@slowromancap\romannumeral #1@}
\makeatother
\usepackage{setspace}

\usepackage{color}

\begin{document}

\title{Spectator electron effects on the two-electron one-photon radiative transition process of double K hole state of Xe$^{q+}$ ion($47\!\leq\!q\!\leq\!52$)}

\author{Xiaobin Ding$^{1}$}
\email{dingxb@nwnu.edu.cn}
\author{Cunqiang Wu$^{1}$}
\author{Denghong Zhang$^{1}$}
\author{Mingwu Zhang$^{2}$}
\author{Yingli Xue$^{2}$}
\author{Deyang Yu$^{2}$}
\author{Chenzhong Dong$^{1}$}

\affiliation{$^1$Key Laboratory of Atomic and Molecular
Physics and Functional Materials of Gansu Province,
College of Physics and Electronic Engineering,
Northwest Normal University, Lanzhou 730070, P. R. China\\
$^2$ Institute of Modern Physics, Chinese Academy of Sciences, Lanzhou 730000, China}
\date{\today}

\begin{abstract}
The double electron E1 transition energies, probabilities, and oscillator strengths between the $2s2p^{n}$ and $1s^{2}2p^{n-1}$($1\!\leq\!n\leq\!6$) configurations of Xe$^{q+}$($47\!\leq\!q\leq\!52$) ions with different spectator electrons have been calculated based on the multi-configuration Dirac-Hartree-Fock method. A reasonable electron correlation model is constructed with the aid of the active space method. The finite mass of nuclear, Breit interaction and QED effects have also been included. The calculated results are in good agreement with the available data. The theoretical spectra of different spectator electrons of the double K hole state have been predicted. The spectator electron effects on the transition spectra have been analyzed in detail. The present results will be helpful for analyzing the high energy X-ray spectrum observed from the interaction between high energy highly charged ions and the surface.
\end{abstract}

\pacs{31.15.vj, 31.30.J–, 32.80.Aa} \maketitle

\section{INTRODUCTION}

The structure and properties of inner shell double-hole atoms (hollow atoms) are essential in modern atomic physics\cite{Briand1976,PhysRevLett.36.164,Safronova1977}. The double hole state in the inner shell is a particular excited state in which an inner shell orbital has no electron occupation, and electrons have occupied the outer orbital. These states can either be created in collisional excitation/ionization of the inner shells of atoms or ions by atom-ion collision\cite{Urrutia2005} and in inner-shell photoionization or photoexcitation processes with a high-energy photon\cite{PhysRevA.79.032708}. They can also be produced by slow highly charged ions interacting with the metal surface in which the electron of the target atom might be captured into the outer shell of the projectile\cite{Briand1996}. These exotic states are unstable and can either decay by emission of Auger electron or high energy X-ray photon, where the former is preferable to the latter for light atom\cite{Chen1991a,article}. However, X-ray spectra of double hole-state of heavy atoms have been observed in high-energy ion atom collisions\cite{Tawara2011,Bliman1989,Volpp1979}, synchrotron radiation\cite{Polasik2011,PhysRevA.34.5168,Kavcic2009}, laser-generated plasma\cite{Elton2000,Lunney1983}, Tokamak\cite{PhysRevA.29.661}, and beam foil spectra\cite{Andriamonje1991}. The study of double hole atoms helps us obtain the structure information of the electrons in this exotic atoms and reveal the influence of electron correlation effect on them. It also provides important diagnostic information for celestial bodies and laboratory plasmas\cite{Porquet2010,Decaux1997}. Hollow atoms also are of great importance for studies of ultrafast dynamics in atoms far from equilibrium and have possible wide-ranging applications in physics, chemistry, biology, and materials science\cite{Costa2006a}.

The radiative deexcitation of an atom with an initially empty K shell may take place either through the more probable one-electron one-photon (OEOP) transition or through the competing weak two-electron one-photon (TEOP) transition. The OEOP transition is the most important mechanism for radiative transition, while the TEOP transition is strictly forbidden in single-particle approximation. TEOP is caused by electron correlation effects. The TEOP transition process was first predicted theoretically by Heisenberg in 1925\cite{Heisenberg1925} and was observed from the ion-atom collision experiments between Ni-Ni,Ni-Fe,Fe-Ni, and Fe-Fe by W$\ddot{o}$lfli et al.in 1975\cite{PhysRevLett.35.656}. Although the TEOP process is weaker than OEOP (about one in a thousand), it is important to understand the high energy X-ray spectrum of double K-shell hole states correctly and determine some important observable quantities, such as lifetime, fluorescence yield and ionization cross-section of these exotic atoms accurately\cite{Ding2020, koziol2017theoretical}.

In the past few decades, many works have been done on the energy levels and transition properties of hole state atoms in the inner shells\cite{Costa2006,safronova1981influence,Martins2004,santos2003two,saha2009effect,Natarajan2008,aaberg1976origin}. R. Kadrekar et al. evaluated the TEOP to OEOP branch ratio of He-like systems of $4\!\leq\!Z\leq\!26$ ions using the multi-configuration Dirac-Fock (MCDF) method, and they found the contribution from TEOP transition is appreciable for light ions\cite{kadrekar2011radiative}. K. Koziol et al. measured TEOP transitions of low-Z atoms by synchrotron radiation\cite{koziol2017theoretical}. They found the discrepancies between experiment and theoretical predictions of the relative intensities of TEOP and shown that the double photoionization perturbation causes these discrepancies. C. Shao et al. observed the X-ray spectra of K-shell hollow krypton atoms produced in single collisions with 52 MeV/u \textrm{-} 197 MeV/u Xe$^{54+}$ ions by heavy-ion storage ring equipped with an internal gas-jet target\cite{Shao2017}. {\L}. Jabłoński et al. firstly observed TEOP X-ray transition in the collision of highly charged Xe$^{26+}$ ions with the surface of the metal Be\cite{Jablonski2020a}.

The hole states created by either light or ion-atom collisions with atom will result in an inner-shell excited state with many electrons occupied in the outer shell. Thus the electron correlation effect is extremely complex.  If the bare ion is used to bombard the target, it has a probability to capture the different number of electrons to the outer shell, which will create a relatively simple double-K hole state with finite spectator electrons. The spectator electron will affect the electron correlation in the double-K hole state and their transition properties. To understand the electron correlation effects and the effects of spectator electrons on the spectra, the structure and radiative properties of a double K-shell hole state with different spectator electrons should be explored. In this work, the energy level structure and radiative transition properties of Xe$^{q+}$ ($47\!\leq\!q\!\leq\!52$) ions have been studied by the multi-configuration Dirac-Hartree-Fock (MCHDF) method combining with the active space method to include the electron correlations effects efficiently. The Breit interaction and quantum electrodynamics (QED) contribution were included perturbatively in separated relativistic configuration interaction (RCI) calculation. The theoretical spectrum of the double K hole state with different spectator electrons is predicted. The present results will be helpful for future theoretical and experimental work.

\section{Theoretical method}
The MCDHF method is widely used to investigate the relativistic, electron correlation, Breit interaction, and QED effects on the complex atoms’ structure and transitions properties\cite{Ding2011,Ding2017a,Aggarwal2016187,PhysRevA.84.062506,natarajan2013two}. I. Grant described the method in detail in his monograph\cite{Grant2007}. The GRASP family codes are developed based on the MCDHF theory in the past 40 years\cite{Grant1980,McKenzie1980,K.1989,GRASP92,Jonsson2007,Joensson2013a,Fischer2019}. The present calculations have been performed using Grasp2K code\cite{Joensson2013a}. Here, only a brief introduction about the MCDHF method was given below.

In the MCDHF method, the atomic state wave function(ASFs) $\psi(PJM_{J})$ for a given state with certain parity P, total angular momentum J, and its z component M$_{J}$ are represented by a linear combination of configuration state functions(CSFs) $\Phi$($\gamma$$_{i}$$PJM_{J}$) with the same P, J, M$_{J}$, which can be expressed as:
\begin{eqnarray}
\psi(PJM_{J})=\sum_{\nu=1}^{N_{c}}c_{i}\Phi(\gamma_{i}PJM_{J}).
\end{eqnarray}
where $N_{c}$ is the number of CSFs, $\gamma_{i}$ denotes all the other quantum numbers necessary to define the configuration, $c_{i}$ is the mixing coefficient for state $i$. The CSFs are the linear combinations of Slater determinants of the many-particle system consisting of single-electron orbital wave functions. The extended optimal level mode is used in the self-consistent field calculations to obtain radial wave functions and the mixing coefficients. The Breit interaction and QED effects, such as vacuum polarization and self-energy, were taken into account perturbatively in a separated relativistic configuration interaction calculation.

After the atomic state function is obtained, the transition probability corresponding to the radiative transitions from the initial state $|\psi_{i}(PJM)\rangle$ to the final state $|\psi_{f}(PJM)\rangle$ can be given by the following equation

\begin{equation}
A_{if}=\frac{4e^{2}\omega_{ij}^{3}}{3\hbar c^{2}}|\langle\psi _{f}(PJM)|Q_{L}^{M}|\psi_{i}(PJM)\rangle|^{2}
\end{equation}
Q$_{L}^{M}$ is the multipole radiation field operator, and $\omega_{ij}$ is the frequency of the photon.

One of the critical tasks in MCDHF calculation is to construct an appropriate CSFs to take the electron correlation effects into account efficiently. At the start, a single-configuration Diarc-Hartree-Fock (DF) calculation was performed to obtain the initial orbitals. Then a systematical expansion of the reference configuration was considered by single and double electron (SD) excitation from the reference configuration to the active orbitals. The optimized wave function is used to calculate the energy level and transition parameters. The  \{1s,2s,2p\} orbitals are treated as the spectroscopic orbitals, and all active orbitals with $n\geq3$ are regarded as correlated orbitals.  Then, the active space was extended to the first layer, i.e., n=3, l=2 ({n3l2}) virtual orbitals, and all the newly added correlation orbitals were optimized with the previous core orbital frozen. With the increase of the active space, the number of CSFs also increased rapidly. To keep the calculation manageable, we only extend the active space orbitals to the principal quantum number $n=6$ and angular quantum number $l=3$. According to the further calculation, the higher angular momentum orbital (l$\geq4$) contribution to the total energy is less than 1\%, so the present calculation was restricted within n$\leq6$,l$\leq3$.

\section{Result and discussion}

\begin{table*}
\caption {The energy eigenvalue (in Hartree) of the excited state $1s^{2}2p^{n}$($2\!\leq\!n\leq\!4$) of Xe$^{50+}$, Xe$^{49+}$ and Xe$^{48+}$ ions. DF denotes the Dirac-Fock calculation, \{n$alb$\} represent the electron correlation effects with active set consists of all orbitals from n=$a$, $l=b$. E$_{R}$ (in eV) is the excitation energy relative to the ground state with the relativistic effects, and E$_{RCI}$ (in eV) represents the excitation energy including the Breit interaction and the QED effect.}
\label{Tab1}
\setlength{\tabcolsep}{8.0mm}{
\begin{tabular}{cccccccccc}
\midrule
\hline
\hline
\multirow{1}{*}{Active sets}                                                      \\\hline
\multicolumn{6}{c}{Be-like Xe ($1s^{2}2p^{2}$)}                  \\
\cmidrule(lr){5-6}
            & $^{1}$S$_{0}$        &  $ ^{1}$D$_{2}$       &$^{3}$P$_{2}$     & $^{3}$P$_{1}$     & $^{3}$P$_{0}$         \\
\midrule
   {DF}     & -3677.77         & -3679.28         & -3692.60      & -3693.31       & -3706.08         \\	
   {n2l1}    & -3677.62         & -3679.28         & -3692.61      & -3693.31       & -3705.88          \\	
   {n3l2}    & -3677.64         & -3679.29        & -3692.62      & -3693.32       & -3705.89         \\	
   {n4l3}    & -3677.64        & -3679.29         & -3692.62 	  & -3693.32       & -3705.89         \\	
   {n5l3}    & -3677.69         & -3679.33         & -3692.67      & -3693.35       & -3705.94        \\	
   {n6l3}    & -3677.72        & -3679.36          & -3692.69      & -3693.38       & -3705.97  \\
   {n7l3}    & -3677.73         & -3679.37         & -3692.70      & -3693.39       & -3705.97        \\
                                                                                                       \\
    E$_{R}$(eV)    &  1070.28      &  1025.72      &  662.83       &  644.20        &  301.68           \\
    E$_{RCI}$(eV)   &  1061.70      &  1013.28      &  654.52       &  639.01        &  302.46   \\
    $Ref^{a}$ &  1071.44      &  1026.52      &  663.96       &  644.74        &  302.62         \\
    $Ref^{a}$ & 1060.35       &1011.56        & 653.68        & 637.63         & 302.03        \\
    $NIST^{b}$  &              &                &653.56       &638.16          &301.70 \\
~\\
\multicolumn{6}{c}{B-like Xe  ($1s^{2}2p^{3}$)}                                                        \\
\cmidrule(lr){5-6}
             & $^{2}$P$_{\frac{3}{2}}$ & $^{2}$P$_{\frac{1}{2}}$ & $^{2}$D$_{\frac{5}{2}}$ & $^{2}$D$_{\frac{3}{2}}$& $^{4}$S$_{\frac{3}{2}}$   \\
\midrule
    {DF}     & -4000.14      & -4015.67      & -4016.77      & -4017.66       & -4030.01               \\	
   {n2l1}    & -4003.07      & -4015.52      & -4016.75      & -4017.65       & -4029.81                     \\	
   {n3l2}    & -4003.11      & -4015.59      & -4016.82      & -4017.70       & -4029.88            \\	
   {n4l3}    & -4033.20      & -4015.68      & -4016.91 	 & -4017.77       & -4029.96             \\	
   {n5l3}    & -4003.21      & -4015.70      & -4016.92      & -4017.79       & -4029.98        \\	
   {n6l3}    & -4003.22      & -4015.71      & -4016.93      & -4017.79       & -4029.99                 \\
                                                                                                \\
    E$_{R}$(eV)    &  1463.21      &  1123.50      &  1090.24      &  1066.69       &  734.87               \\
    E$_{RCI}$(eV)   &  1443.63      &  1110.34      &  1071.97      &  1052.25       &  725.36       \\
   $Ref^{c}$ &  1064.46      &  1125.38      &  1091.81      &  1067.76       &  736.65              \\
   $Ref^{d}$ &  1443.91      &  1111.59      &  1073.12      &  1052.69       &  726.89              \\
~\\
\multicolumn{6}{c}{C-like Xe ($1s^{2}2p^{4}$)}                                                                     \\
\cmidrule(lr){5-6}
            & $^{1}$S$_{0}$ & $^{1}$D$_{2}$ & $^{3}$P$_{2}$& $^{3}$P$_{1}$  & $^{3}$P$_{0}$               \\
\midrule
     {DF}     & -4318.19      & -4331.55    & -4345.06    & -4332.28      & -4343.66                     \\	
   {n2l1}    & -4317.96      & -4331.37     & -4344.77    & -4332.12      & -4343.25                            \\	
   {n3l2}    & -4318.04      & -4331.49     & -4344.91    & -4332.21      & -4343.40                   \\	
   {n4l3}    & -4318.23      & -4331.67     & -4345.08     & -4332.39 	  & -4343.58                  \\	
   {n5l3}    & -4318.27      & -4331.72     & -4345.12     & -4332.43      & -4343.62               \\	
   {n6l3}    & -4318.28      & -4331.73     & -4345.13     & -4332.44      & -4343.64                       \\
                                                                                                   \\
    E$_{R}$(eV)     &  1875.94        &  1510.05       &  1145.24       &  1490.70      &  1185.98                     \\
    E$_{RCI}$(eV)   &  1855.72        &  1491.06       &  1132.34       &  1474.66      &  1176.54           \\
   $Ref^{c}$        &  1877.59        &  1512.10       &  1147.40       &  1492.32      &  1188.59       \\
   $Ref^{a}$        &  1852.61        &  1488.86       & 1130.41        &  1471.85      &  1174.88      \\
~\\
\hline
\hline
\end{tabular}}
\begin{tablenotes}
\item[] $^{\rm a}$ Reference\cite{Cheng1979}.
\item[] $^{\rm b}$ Reference\cite{NIST_ASD}.
\item[] $^{\rm c}$ Reference\cite{aggarwal2010energy}.
\item[] $^{\rm d}$ Reference\cite{{0953-4075-43-3-035005}}\\
\end{tablenotes}
\end{table*}

The energy eigenvalues and the excitation energy of the ground configuration 1s$^{2}$2p$^{2}$, 1s$^{2}$2p$^{3}$, and 1s$^{2}$2p$^{4}$  of Be-, B- and C-like Xe ions are given in Table \ref{Tab1} with available experimental and theoretical data for comparison. The excitation energy E$_{R}$ was calculated against the corresponding ground state 1s$^{2}$2s$^{2}$2p$^{n}$($0\!\leq\!n\!\leq\!2$) of each ion with relativistic effects. E$_{RCI}$ represents the excitation energy with  Breit interaction and QED effect. It is found that the energy eigenvalues converged with the increase of active space. The contribution from Breit interaction and QED effect to the energy level is about 1.3\%, indicating that these contributions are important and cannot be ignored. The excitation energy is in good agreement with the available results within 0.3\%, which means that the electron correlation model constructed in the present work is reasonable.

\begin{table*}[]
\caption{The two-electron one-photon (TEOP) and one-electron one-photon (OEOP) transition energy (in eV) and the $1s2s$-$1s^{2}$ energy level difference (in eV) of He-like Xe ion.}
\label{table2}
\setlength{\tabcolsep}{9.0mm}{
\begin{tabular}{cccccccccc}
\midrule
\hline
\hline
\multicolumn{5}{c}{He-like Xe}                                                                \\\hline
\multirow{1}{*}{Confuguration}  & \multicolumn{1}{c}{Transtion} & \multicolumn{1}{c}{Energy}   & \multicolumn{1}{c}{Type}    \\
\cmidrule(lr){1-2}
\midrule
\multirow{2}{*}{$2s2p$-$1s^{2}$}                &   $^{3}$P$_{1}$-$^{1}$S$_{0}$     &           60925.53     &       TEOP    \\
                                                & $^{1}$P$_{1}$-$^{1}$S$_{0}$       &            61385.87      &       TEOP    \\
\multirow{2}{*}{$2s2p$-$1s2s$}                   &   $^{3}$P$_{1}$-$^{1}$S$_{0}$              &30712.78       &     OEOP    \\

                                                 &   $^{1}$P$_{1}$-$^{3}$S$_{1}$        &   31259.35        &      OEOP   \\
\multicolumn{5}{c}{Excitation Energy}       \\
\multirow{1}{*}{}&    \multicolumn{1}{c}{}& \multicolumn{1}{c}{Present} & \multicolumn{1}{c}{NIST\cite{NIST_ASD}}   & \multicolumn{1}{c}{Ref\cite{plante1994relativistic}}    \\
\multirow{2}{*}{ $1s2s$-$1s^{2}$ }                         &$^{1}$S$_{0}$-$^{1}$S$_{0}$         &  30212.75         & 30213.86    &30212.44       \\
                                                           &$^{3}$S$_{1}$-$^{1}$S$_{0}$            &  30126.52       &30128.77       &30127.36                          \\

~\\
\hline
\hline
\end{tabular}}
\end{table*}

The transition energies of the TEOP and OEOP of He-like Xe ions are shown in TABLE \ref{table2}. To the best of the authors' knowledge, there are no calculations and experiments on the TEOP and OEOP transitions for Xe$^{q+}$ ($47\!\leq\!q\!\leq\!52$) ions. To verify the validity of our calculation results, the excitation energy of $1s2s$ of He-like Xe was calculated using the difference between the TEOP and OEOP transition energies. Compared with the available results\cite{plante1994relativistic} and NIST database\cite{NIST_ASD}, about 0.01\% agreement was obtained. In such heavy and highly charged Xe ions, the electronic correlation models could be extended safely to the highly excited state 2s2p$^{n}$($1\!\leq\!n\leq\!6$) of Xe ions. According to the present calculation, the eigenvalues of these highly excited states also tend to converge with the increase of the active space.

\begin{table*}[]
\caption{Some selected two-electron one-photon transition energy (eV), probability $A_{ij}$ ($s^{-1}$) and oscillator strength gf between $2s2p^{n}$-$1s^{2}2p^{n-1}$ ($1\!\leq\!n\leq\!6$) of Xe$^{q+}$ ion($47\!\leq\!q\!\leq\!52$).}
\label{Tab3}
\setlength{\tabcolsep}{8.0mm}{
\begin{tabular}{ccccccccc}
\midrule
\hline
\hline
\multirow{1}{*}{Configuration} & \multicolumn{1}{c}{Transition} &   \multicolumn{1}{c}{Energy(eV)}      &   \multicolumn{1}{c}{A$_{ij}$(s$^{-1}$)} &   \multicolumn{1}{c}{gf}                          \\\hline
\multicolumn{5}{c}{He-like Xe}                                                               \\
\multirow{2}{*}{$2s2p$-1s$^{2}$}                 & $^{3}$P$_{1}$-$^{1}$S$_{0}$         &60925.53                & 1.3775E+11          &5.8675E-06                \\
                                                 & $^{1}$P$_{1}$-$^{1}$S$_{0}$         &61385.87                &3.1980E+11          &2.5656E-06                 \\
\multicolumn{5}{c}{Li-like Xe}                                                                                \\
\multirow{8}{*}{$2s2p^{2}$-1s$^{2}2p$}           & $^{2}$S$_{\frac{1}{2}}$-$^{2}$P$_{\frac{1}{2}}$         &60679.27                & 1.4240E+11          & 1.7825E-06   \\
                                                 & $^{4}$P$_{\frac{5}{2}}$-$^{2}$P$_{\frac{3}{2}}$         &60698.29                & 1.0757E+11          & 4.0371E-06   \\
                                                 & $^{2}$P$_{\frac{1}{2}}$-$^{2}$P$_{\frac{3}{2}}$         &60748.74                & 9.9012E+10          & 1.2366E-06    \\
                                                 & $^{2}$D$_{\frac{3}{2}}$-$^{2}$P$_{\frac{3}{2}}$         &60749.44                & 6.5676E+10          & 1.6405E-06     \\
                                                 & $^{2}$P$_{\frac{1}{2}}$-$^{2}$P$_{\frac{1}{2}}$         &61122.67                & 1.1047E+11          & 1.3629E-06     \\
                                                 & $^{2}$D$_{\frac{3}{2}}$-$^{2}$P$_{\frac{1}{2}}$         &61123.37                & 8.2374E+10          & 2.0325E-06     \\
                                                 & $^{4}$P$_{\frac{1}{1}}$-$^{2}$P$_{\frac{3}{2}}$         &61172.03                & 7.4831E+10          & 9.2171E-07     \\
                                                 & $^{4}$P$_{\frac{3}{2}}$-$^{2}$P$_{\frac{3}{2}}$         &61172.03                & 1.4413E+11          & 3.5506E-06   \\
\multicolumn{5}{c}{Be-like Xe}                                                                                 \\
\multirow{16}{*}{$2s2p^{3}$-1s$^{2}2p^{2}$}      & $^{3}$P$_{2}$-$^{3}$P$_{2}$         &60448.39                & 9.8652E+10          & 3.1110E-06   \\
                                                 & $^{3}$P$_{2}$-$^{3}$P$_{1}$         &60463.87                & 1.2648E+10          & 3.9864E-06   \\
                                                 & $^{3}$D$_{3}$-$^{1}$D$_{2}$         &60493.00                & 1.7797E+10          & 7.8454E-06   \\
                                                 & $^{3}$P$_{1}$-$^{3}$P$_{2}$         &60510.93                & 1.7567E+10          & 3.3170E-06   \\
                                                 & $^{3}$S$_{1}$-$^{1}$S$_{0}$         &60512.51                & 1.5159E+11          & 3.9864E-06   \\
                                                 & $^{3}$P$_{1}$-$^{1}$D$_{2}$         &60546.70                & 1.1555E+11          & 3.9864E-06   \\
                                                 & $^{3}$D$_{2}$-$^{1}$D$_{2}$         &60557.22                & 1.1092E+11          & 3.4853E-06   \\
                                                 & $^{3}$D$_{2}$-$^{3}$P$_{2}$         &60851.81 	            & 9.5912E+10	      & 4.1784E-06    \\
                                                 & $^{1}$P$_{1}$-$^{3}$P$_{0}$         &60863.03 	            & 3.3388E+11	      & 6.2315E-06   \\
                                                 & $^{3}$P$_{0}$-$^{3}$P$_{1}$         &60906.49 	            & 2.5400E+11	      & 1.5780E-06   \\
                                                 & $^{3}$D$_{2}$-$^{3}$P$_{2}$         &60916.03  	            & 4.7964E+11	      & 1.4894E-05   \\
                                                 & $^{5}$S$_{2}$-$^{1}$D$_{2}$         &60916.49  	            & 3.0770E+11	      & 9.5545E-06   \\
                                                 & $^{3}$S$_{1}$-$^{3}$P$_{2}$         &60919.68  	            & 3.2813E+11	      & 6.1129E-06   \\
                                                 & $^{3}$P$_{1}$-$^{3}$P$_{1}$         &60921.00  	            & 4.7527E+11	      & 8.8536E-06  \\
                                                 & $^{3}$D$_{1}$-$^{1}$S$_{0}$         &60927.02  	            & 1.9703E+11	      & 3.6696E-06  \\
                                                 & $^{3}$D$_{1}$-$^{1}$D$_{2}$         &60975.37  	            & 4.5540E+11	       &8.4682E-06  \\
 \multicolumn{5}{c}{B-like Xe}                                                                                     \\
\multirow{13}{*}{$2s2p^{4}$-1s$^{2}2p^{3}$}      & $^{2}$D$_{\frac{5}{2}}$-$^{2}$D$_{\frac{3}{2}}$         &60260.81              & 9.0883E+10          & 3.4606E-06   \\
                                                 & $^{4}$P$_{\frac{1}{2}}$-$^{2}$P$_{\frac{1}{2}}$         &60272.15              & 1.4583E+11          & 1.8503E-06   \\
                                                 & $^{4}$P$_{\frac{5}{2}}$-$^{2}$P$_{\frac{3}{2}}$         &60296.76               & 1.1495E+11          & 4.3717E-06   \\
                                                 & $^{2}$D$_{\frac{3}{2}}$-$^{2}$D$_{\frac{5}{2}}$         &60311.27              & 1.0261E+11            & 2.6003E-06   \\
                                                 & $^{2}$P$_{\frac{3}{2}}$-$^{2}$P$_{\frac{3}{2}}$         &60343.67             & 8.9080E+10          & 2.2551E-06   \\
                                                 & $^{4}$P$_{\frac{3}{2}}$-$^{2}$D$_{\frac{3}{2}}$         &60652.92              & 1.6728E+11        & 4.1912E-06   \\
                                                 & $^{4}$P$_{\frac{1}{2}}$-$^{4}$S$_{\frac{3}{2}}$         &60657.13              & 1.4585E+11          & 1.8271E-06   \\
                                                 & $^{2}$D$_{\frac{3}{2}}$-$^{4}$S$_{\frac{3}{2}}$         &60657.88  	          & 2.8186E+11	       & 7.0618E-06    \\
                                                 & $^{2}$P$_{\frac{1}{2}}$-$^{2}$P$_{\frac{1}{2}}$         &60665.98  	             & 1.2444E+11	       & 1.5585E-06  \\
                                                 & $^{4}$P$_{\frac{5}{2}}$-$^{2}$D$_{\frac{5}{2}}$         &60668.41 	             & 1.5660E+11	       & 5.8830E-06  \\
                                                 & $^{2}$P$_{\frac{3}{2}}$-$^{2}$D$_{\frac{5}{2}}$         &60715.32  	             & 2.5395E+11	       & 6.3504E-06 \\
                                                 & $^{1}$S$_{\frac{1}{2}}$-$^{2}$P$_{\frac{3}{2}}$         &60723.99  	             & 3.4008E+11	       & 4.2508E-06   \\
                                                 & $^{2}$P$_{\frac{1}{2}}$-$^{2}$D$_{\frac{3}{2}}$         &60724.07   	             & 2.2823E+11	       & 2.8529E-06    \\
 \multicolumn{5}{c}{C-like Xe}                                                                \\
\multirow{8}{*}{$2s2p^{5}$-1s$^{2}2p^{4}$}       & $^{3}$P$_{2}$-$^{1}$D$_{2}$         &60055.97                 & 1.1975E+11          & 3.8259E-06   \\
                                                 & $^{3}$P$_{2}$-$^{3}$P$_{1}$         &60072.38                 & 1.3351E+11          & 4.2630E-06   \\
                                                 & $^{3}$P$_{1}$-$^{1}$D$_{2}$         &60110.63                 & 1.7244E+11          & 3.2994E-06   \\
                                                 & $^{1}$P$_{1}$-$^{1}$S$_{0}$         &60116.59                 & 1.7946E+11          & 3.3170E-06   \\
                                                 & $^{3}$P$_{2}$-$^{3}$P$_{2}$         &60414.71                 & 2.5805E+11          & 8.1467E-06   \\
                                                 & $^{3}$P$_{1}$-$^{3}$P$_{2}$         &60469.37                 & 3.7752E+11          & 7.1380E-06   \\
                                                 & $^{3}$P$_{0}$-$^{3}$P$_{1}$         &60465.68                 & 4.9222E+11          & 3.1026E-06  \\
                                                 & $^{1}$P$_{1}$-$^{1}$D$_{2}$         &60481.27  	             & 4.7273E+11	       & 8.9347E-06    \\
\multicolumn{5}{c}{N-like Xe}                                                             \\
\multirow{2}{*}{$2s2p^{6}$-1s$^{2}2p^{5}$}       & $^{2}$S$_{\frac{1}{2}}$-$^{2}$P$_{\frac{1}{2}}$         &59894.57                &2.5963E+11          & 3.3359E-06   \\
                                                 & $^{2}$S$_{\frac{1}{2}}$-$^{2}$P$_{\frac{3}{2}}$         &60238.05                &4.6951E+11          & 5.9638E-06   \\
\hline
\end{tabular}}
\end{table*}

The TEOP transition energy, probability and oscillator strength with relatively large transition probability of Xe$^{q+}$ ($47\!\leq\!q\!\leq\!52$) ion are presented in TABLE \ref{Tab3}. In the relativistic atomic structure calculation, the consistency of the transition probability calculated from different gauge (i.e., Babushkin and Coulomb) could show the accuracy of the wavefunction\cite{Cowan1981,Elsayed2015}. It was found in the present calculation that the ratio of the transition probability from different gauge is about $1.19\sim1.70$ from He-like to N-like Xe ions, this also indicates that the wavefunction used in the present calculation was accurate enough. The most important correlation effects were included in the present calculation. For brevity, only transition probability and oscillator strength in Babushkin gauge, which corresponding to the length gauge in the non-relativistic limit, are given in TABLE ~\ref{Tab3}.

\begin{figure*}[!h]
\begin{minipage}[t]{1\linewidth}
\centering
\includegraphics[width=17.0cm,height=12.0cm]{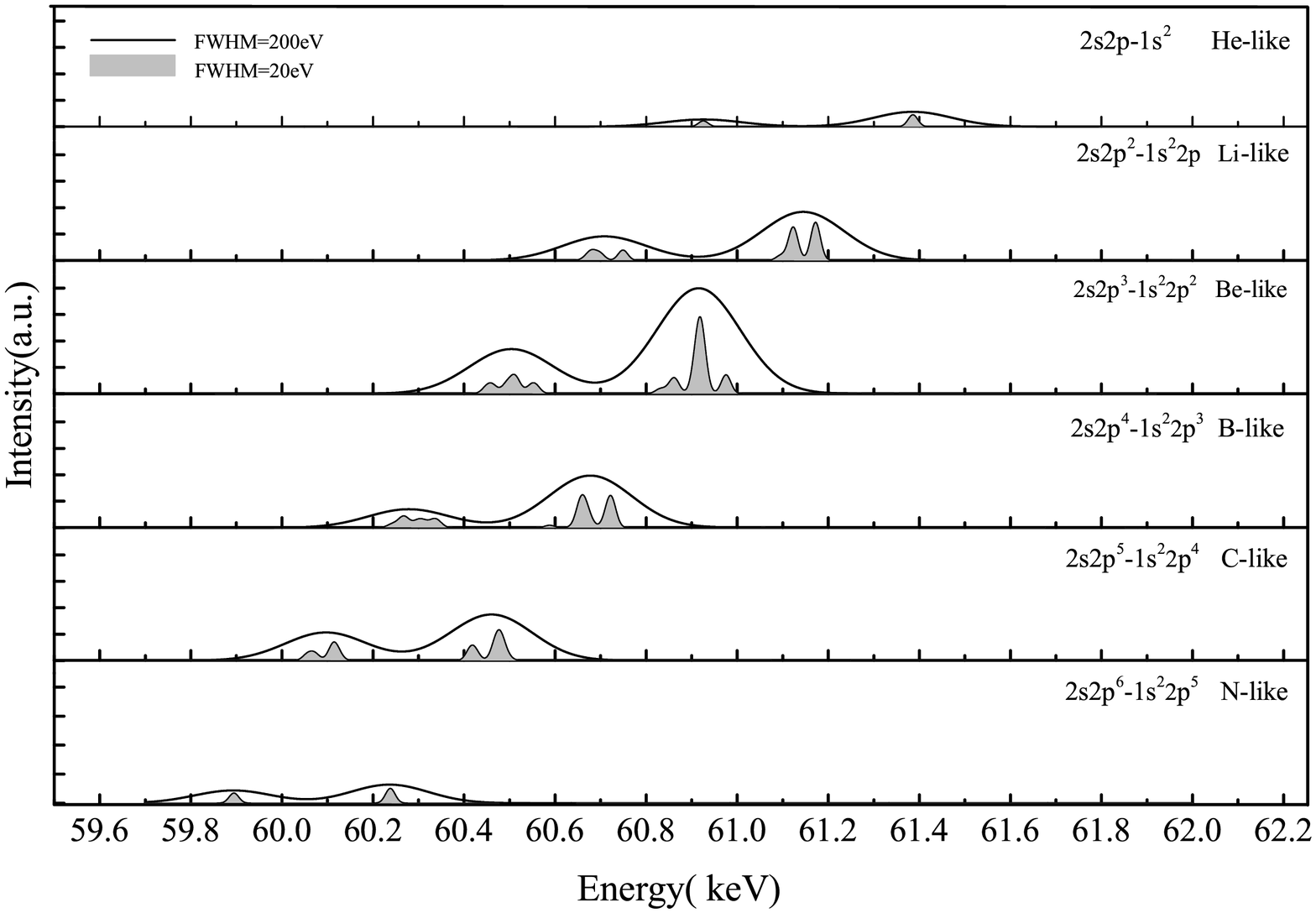}\label{Fig.1}
\caption{Relative intensity of the two-electron one-photon transition  $2s2p^{n}$$\rightarrow$$1s^{2}2p^{n-1}$($1\!\leq\!n\leq\!6$) of Xe ions. The curve is calculated by the transition probability with full width at half maximum (FWHM) 200 eV which is the typical resolution of the X-ray spectrum around 60 KeV, while the shadow is obtained with FWHM=20eV.}\label{fig1}
\end{minipage}
\end{figure*}

Fig.~\ref{fig1} shows that the relative intensity of the TEOP transition from double-K hole states of $2s2p^{n}$ to $1s^{2}2p^{n-1}$($1\!\leq\!n\leq\!6$) Xe$^{q+}$ ions. The transition probability is proportional to the intensity of the transition line. Each individual transition was assumed to have a Gaussian profile to consider the natural, collisional, and Doppler broadening effects. At present, the best resolution of high energy X-ray spectra around 60 keV is about 200eV to 500eV\cite{uhlig2015high}. The predicted spectra with the full width at half maximum (FWHM) 200eV are given in Fig.~\ref{fig1}. All the spectra are consist of two peaks, which corresponds to the diagram transition from the $^{1}P$ and $^{3}P$ parent term of $2s2p$ to the $^{1}$S of $1s^{2}$, respectively. The transition peak gets shifted towards the low energy side with the increased number of spectator electrons. The "red-shift" energy per $2p$ spectator electron is about 240eV. However, with the development of spectroscopy technology, the resolution is expected to be further improved. The shadow of each figure gives the spectrum with FWHM at 20eV. The detailed structure of these peaks could be studied with higher accuracy in the future.

\begin{figure*}[!h]
	\begin{minipage}[t]{1\linewidth}
		\centering
		\includegraphics[width=18.5cm,height=12.5cm]{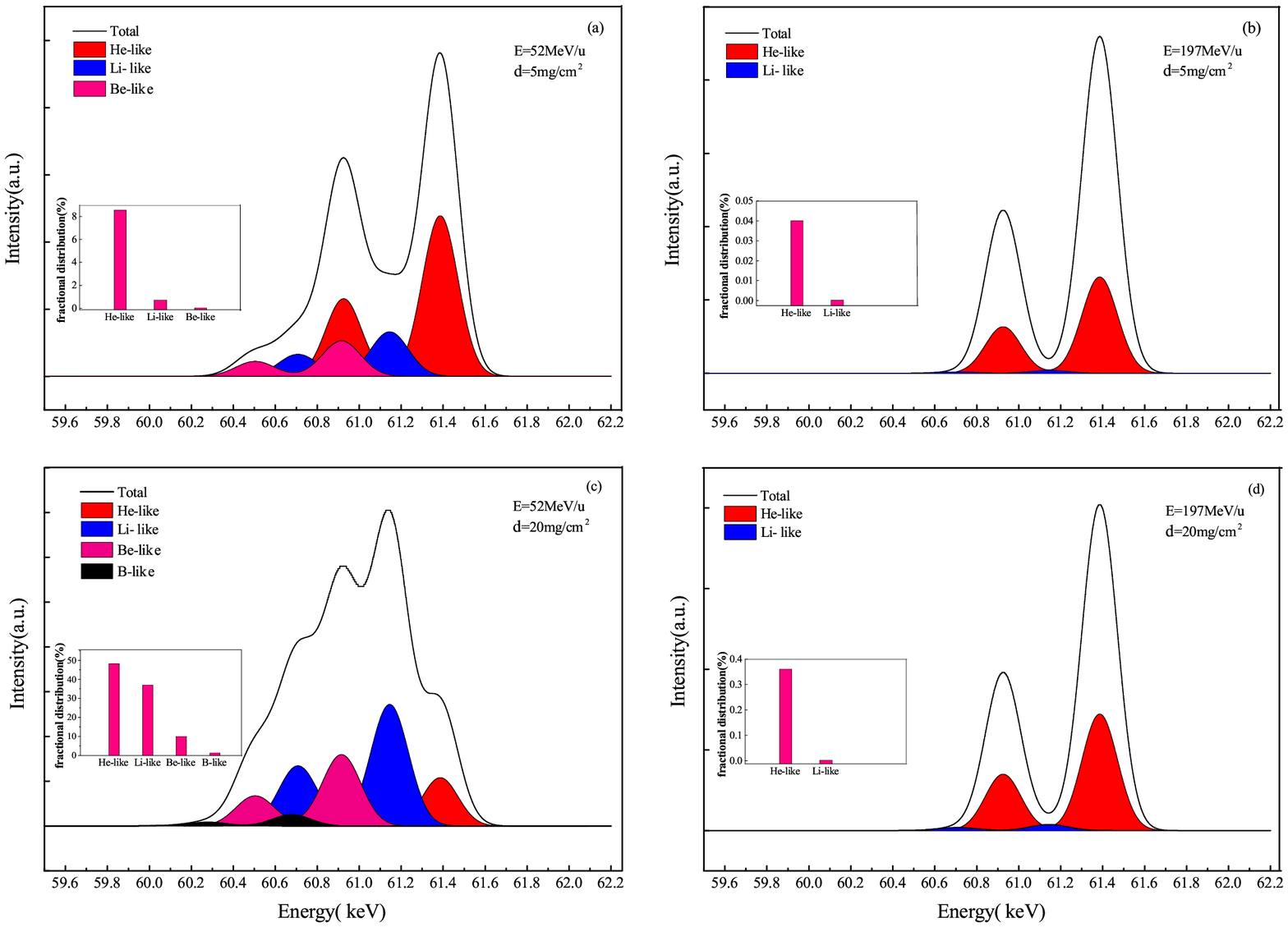}\label{Fig.2}
		\caption{The possible two-electron one-photon X-ray spectrum from $2s2p^{n}$$\rightarrow$$1s^{2}2p^{n-1}$($1\!\leq\!n\leq\!6$) under different projectile energy (E) of the bare Xe nucleus and target thickness (d). (a) The thickness of the target is 5mg/cm$^{2}$ and the projectile energy is 52MeV/u; (b) The thickness of the target is 5mg/cm$^{2}$ and the projectile energy is 197MeV/u; (c) The thickness of the target is 20mg/cm$^{2} $ and the projectile energy is 52MeV/u; (d) The thickness of the target is 20mg/cm$^{2}$ and the projectile energy is 197MeV/u.}\label{fig2}
	\end{minipage}
\end{figure*}

Using highly charged, even bare ion bombarded to the target is an efficient way to produce double-K hole states with simple structure\cite{briand1990production,winter1999hollow,Shao2017}. The projectile might capture a different number of electrons from the target to the outer shells during the interaction. Generally, the collisional products will in different ionization stages. Therefore, to predict the observed spectrum from mixing double-K hole states, the ions’ population should be taken into account. The charge state distribution after the ion-surface interaction can be predicted using the semi-empirical model within the LISE++ code\cite{Briand_1989,LamourExtension,Vockenhuber2013,rozet1996etacha,tarasov2016lise++,tarasov1980266}. We will propose an experiment to observe the TEOP of Xe$^{q+}$($47\!\leq\!q\!\leq\!52$) ions. The initial double K-hole state can be produced by bare Xe$^{54+}$ ions collided with a carbon target. Therefore, the charge state distribution after the Xe$^{54+}$ ions (with the energy of 52 MeV/u and 197 MeV/u) collide with the carbon targets in different thickness are calculated. The fractional population contribution and corresponding transition spectra in different experiment conditions are obtained with FWHF=200eV as shown in Fig.~\ref{fig2}. It can be found from the figure that except for Fig.~\ref{fig2}(c), the other three transition spectra of (a), (b), and (d) show a relatively simple bimodal structure that mainly corresponds to the He-like ion.

When the thickness of the target is the same, and the incident energy of projectile ions is different, as shown in (a) and (b), the spectral peak structure is similar, but the intensity of the former is about 20 times higher than that of the latter. This caused by the fact that the high energy projectile is too fast to capture electrons from the target. Therefore, only a tiny fractional population was obtained for the high energy projectile. Comparing the Fig.~\ref{fig2}(a) and (c), the peak structure is significantly different, which indicates that the thicker target will have more chances to produce a more highly ionized double K-hole state. It is recommended to use high energy projectiles with thick target to observe the transition wavelength of double K-hole state for the He-like Xe ions.

\section{Conclusion}

Two-electron one-photon (TEOP) E1 transition energies, probabilities, and oscillator strengths between the $2s2p^{n}$ and $1s^{2}2p^{n-1}$ ($1\!\leq\!n\leq\!6$) configurations have been calculated using the Grasp2k package which based on the MCDHF method. Electron correlation effects were handled systematically. The Breit interaction and QED effects further corrected the transition energy. An appropriate electron correlation model for the double K-hole state was constructed. For the configuration of $1s^{2}2p^{n-1}$, it is found that the current calculation results are in good agreement with the previous work. By extending the correlation model for the $2s2p^{n}$ configuration, the TEOP transition from $2s2p^{n}$ to $1s^{2}2p^{n-1}$ ($1\!\leq\!n\leq\!6$) was calculated for different spectator electrons. To the best of authors' knowledge, there are no experimental data available for this type of transitions considered in the present work. The basic atomic data is critical in judging the relative importance of this unusual transition and checking the theoretical predictions on such anomalous level structure of highly charged ions. It is expected to be verified in future experiments.

\section{Acknowledgments}

This work was supported by National Nature Science Foundation of China, Grant No: U1832126, 11874051, National Key Research and Development Program of China, Grant No:2017YFA0402300.\\
\footnotesize
\bibliography{myBIB}

\begin{thebibliography}{64}
\expandafter\ifx\csname natexlab\endcsname\relax\def\natexlab#1{#1}\fi
\expandafter\ifx\csname bibnamefont\endcsname\relax
  \def\bibnamefont#1{#1}\fi
\expandafter\ifx\csname bibfnamefont\endcsname\relax
  \def\bibfnamefont#1{#1}\fi
\expandafter\ifx\csname citenamefont\endcsname\relax
  \def\citenamefont#1{#1}\fi
\expandafter\ifx\csname url\endcsname\relax
  \def\url#1{\texttt{#1}}\fi
\expandafter\ifx\csname urlprefix\endcsname\relax\def\urlprefix{URL }\fi
\providecommand{\bibinfo}[2]{#2}
\providecommand{\eprint}[2][]{\url{#2}}

\bibitem[{\citenamefont{Briand}(1976)}]{Briand1976}
\bibinfo{author}{\bibfnamefont{J.~P.} \bibnamefont{Briand}},
  \bibinfo{journal}{Phys. Rev. Lett.} \textbf{\bibinfo{volume}{37}},
  \bibinfo{pages}{59} (\bibinfo{year}{1976}).

\bibitem[{\citenamefont{Nagel et~al.}(1976)\citenamefont{Nagel, Burkhalter,
  Knudson, and Hill}}]{PhysRevLett.36.164}
\bibinfo{author}{\bibfnamefont{D.~J.} \bibnamefont{Nagel}},
  \bibinfo{author}{\bibfnamefont{P.~G.} \bibnamefont{Burkhalter}},
  \bibinfo{author}{\bibfnamefont{A.~R.} \bibnamefont{Knudson}},
  \bibnamefont{and} \bibinfo{author}{\bibfnamefont{K.~W.} \bibnamefont{Hill}},
  \bibinfo{journal}{Phys. Rev. Lett.} \textbf{\bibinfo{volume}{36}},
  \bibinfo{pages}{164} (\bibinfo{year}{1976}).

\bibitem[{\citenamefont{Safronova and Senashenko}(1977)}]{Safronova1977}
\bibinfo{author}{\bibfnamefont{U.~I.} \bibnamefont{Safronova}}
  \bibnamefont{and} \bibinfo{author}{\bibfnamefont{V.~S.}
  \bibnamefont{Senashenko}}, \bibinfo{journal}{J. Phys. B: At. Mol. Phys.}
  \textbf{\bibinfo{volume}{10}}, \bibinfo{pages}{L271} (\bibinfo{year}{1977}).

\bibitem[{\citenamefont{Urrutia et~al.}(2005)\citenamefont{Urrutia, Artemyev,
  Braun, Brenner, Bruhns, Draganič, Martínez, Lapierre, Mironov, Scofield
  et~al.}}]{Urrutia2005}
\bibinfo{author}{\bibfnamefont{J.~C.~L.} \bibnamefont{Urrutia}},
  \bibinfo{author}{\bibfnamefont{A.}~\bibnamefont{Artemyev}},
  \bibinfo{author}{\bibfnamefont{J.}~\bibnamefont{Braun}},
  \bibinfo{author}{\bibfnamefont{G.}~\bibnamefont{Brenner}},
  \bibinfo{author}{\bibfnamefont{H.}~\bibnamefont{Bruhns}},
  \bibinfo{author}{\bibfnamefont{I.}~\bibnamefont{Draganič}},
  \bibinfo{author}{\bibfnamefont{A.~G.} \bibnamefont{Martínez}},
  \bibinfo{author}{\bibfnamefont{A.}~\bibnamefont{Lapierre}},
  \bibinfo{author}{\bibfnamefont{V.}~\bibnamefont{Mironov}},
  \bibinfo{author}{\bibfnamefont{J.}~\bibnamefont{Scofield}},
  \bibnamefont{et~al.}, \bibinfo{journal}{Nucl. Instrum. Methods Phys. Res.,
  Sect. B} \textbf{\bibinfo{volume}{235}}, \bibinfo{pages}{85 }
  (\bibinfo{year}{2005}).

\bibitem[{\citenamefont{Fennane et~al.}(2009)\citenamefont{Fennane, Dousse,
  Hoszowska, Berset, Cao, Maillard, Szlachetko, Szlachetko, and Kav\ifmmode
  \check{c}\else \v{c}\fi{}i\ifmmode~\check{c}\else
  \v{c}\fi{}}}]{PhysRevA.79.032708}
\bibinfo{author}{\bibfnamefont{K.}~\bibnamefont{Fennane}},
  \bibinfo{author}{\bibfnamefont{J.}~\bibnamefont{Dousse}},
  \bibinfo{author}{\bibfnamefont{J.}~\bibnamefont{Hoszowska}},
  \bibinfo{author}{\bibfnamefont{M.}~\bibnamefont{Berset}},
  \bibinfo{author}{\bibfnamefont{W.}~\bibnamefont{Cao}},
  \bibinfo{author}{\bibfnamefont{Y.}~\bibnamefont{Maillard}},
  \bibinfo{author}{\bibfnamefont{J.}~\bibnamefont{Szlachetko}},
  \bibinfo{author}{\bibfnamefont{M.}~\bibnamefont{Szlachetko}},
  \bibnamefont{and} \bibinfo{author}{\bibfnamefont{M.}~\bibnamefont{Kav\ifmmode
  \check{c}\else \v{c}\fi{}i\ifmmode~\check{c}\else \v{c}\fi{}}},
  \bibinfo{journal}{Phys. Rev. A} \textbf{\bibinfo{volume}{79}},
  \bibinfo{pages}{032708} (\bibinfo{year}{2009}).

\bibitem[{\citenamefont{Briand et~al.}(1996)\citenamefont{Briand, Thuriez,
  Giardino, Borsoni, Froment, Eddrief, and Sebenne}}]{Briand1996}
\bibinfo{author}{\bibfnamefont{J.~P.} \bibnamefont{Briand}},
  \bibinfo{author}{\bibfnamefont{S.}~\bibnamefont{Thuriez}},
  \bibinfo{author}{\bibfnamefont{G.}~\bibnamefont{Giardino}},
  \bibinfo{author}{\bibfnamefont{G.}~\bibnamefont{Borsoni}},
  \bibinfo{author}{\bibfnamefont{M.}~\bibnamefont{Froment}},
  \bibinfo{author}{\bibfnamefont{M.}~\bibnamefont{Eddrief}}, \bibnamefont{and}
  \bibinfo{author}{\bibfnamefont{C.}~\bibnamefont{Sebenne}},
  \bibinfo{journal}{Phys. Rev. Lett.} \textbf{\bibinfo{volume}{77}},
  \bibinfo{pages}{1452} (\bibinfo{year}{1996}).

\bibitem[{\citenamefont{Chen}(1991)}]{Chen1991a}
\bibinfo{author}{\bibfnamefont{M.~H.} \bibnamefont{Chen}},
  \bibinfo{journal}{Phys. Rev. A} \textbf{\bibinfo{volume}{44}},
  \bibinfo{pages}{239} (\bibinfo{year}{1991}).

\bibitem[{\citenamefont{Natarajan and Natarajan}(2008)}]{article}
\bibinfo{author}{\bibfnamefont{A.}~\bibnamefont{Natarajan}} \bibnamefont{and}
  \bibinfo{author}{\bibfnamefont{L.}~\bibnamefont{Natarajan}},
  \bibinfo{journal}{Phys. Rev. A} \textbf{\bibinfo{volume}{109}},
  \bibinfo{pages}{2281} (\bibinfo{year}{2008}).

\bibitem[{\citenamefont{Tawara and Richard}(2011)}]{Tawara2011}
\bibinfo{author}{\bibfnamefont{H.}~\bibnamefont{Tawara}} \bibnamefont{and}
  \bibinfo{author}{\bibfnamefont{P.}~\bibnamefont{Richard}},
  \bibinfo{journal}{Can. J. Phys.} \textbf{\bibinfo{volume}{80}},
  \bibinfo{pages}{1579} (\bibinfo{year}{2011}).

\bibitem[{\citenamefont{Bliman et~al.}(1989)\citenamefont{Bliman, Indelicato,
  Hitz, Marseille, and Desclaux}}]{Bliman1989}
\bibinfo{author}{\bibfnamefont{S.}~\bibnamefont{Bliman}},
  \bibinfo{author}{\bibfnamefont{P.}~\bibnamefont{Indelicato}},
  \bibinfo{author}{\bibfnamefont{D.}~\bibnamefont{Hitz}},
  \bibinfo{author}{\bibfnamefont{P.}~\bibnamefont{Marseille}},
  \bibnamefont{and} \bibinfo{author}{\bibfnamefont{J.}~\bibnamefont{Desclaux}},
  \bibinfo{journal}{J. Phys. B: At., Mol. Opt. Phys.}
  \textbf{\bibinfo{volume}{22}}, \bibinfo{pages}{2741} (\bibinfo{year}{1989}).

\bibitem[{\citenamefont{Volpp et~al.}(1979)\citenamefont{Volpp, Schuch, Nolte,
  and Schmidt-Bocking}}]{Volpp1979}
\bibinfo{author}{\bibfnamefont{J.}~\bibnamefont{Volpp}},
  \bibinfo{author}{\bibfnamefont{R.}~\bibnamefont{Schuch}},
  \bibinfo{author}{\bibfnamefont{G.}~\bibnamefont{Nolte}}, \bibnamefont{and}
  \bibinfo{author}{\bibfnamefont{H.}~\bibnamefont{Schmidt-Bocking}},
  \bibinfo{journal}{J. Phys. B: At. Mol. Phys.} \textbf{\bibinfo{volume}{12}},
  \bibinfo{pages}{L325} (\bibinfo{year}{1979}).

\bibitem[{\citenamefont{Polasik et~al.}(2011)\citenamefont{Polasik,
  Słabkowska, Rzadkiewicz, Kozioł, Starosta, Wiatrowska-Kozioł, Dousse, and
  Hoszowska}}]{Polasik2011}
\bibinfo{author}{\bibfnamefont{M.}~\bibnamefont{Polasik}},
  \bibinfo{author}{\bibfnamefont{K.}~\bibnamefont{Słabkowska}},
  \bibinfo{author}{\bibfnamefont{J.}~\bibnamefont{Rzadkiewicz}},
  \bibinfo{author}{\bibfnamefont{K.}~\bibnamefont{Kozioł}},
  \bibinfo{author}{\bibfnamefont{J.}~\bibnamefont{Starosta}},
  \bibinfo{author}{\bibfnamefont{E.}~\bibnamefont{Wiatrowska-Kozioł}},
  \bibinfo{author}{\bibfnamefont{J.~C.} \bibnamefont{Dousse}},
  \bibnamefont{and}
  \bibinfo{author}{\bibfnamefont{J.}~\bibnamefont{Hoszowska}},
  \bibinfo{journal}{Phys. Rev. Lett.} \textbf{\bibinfo{volume}{107}},
  \bibinfo{pages}{073001} (\bibinfo{year}{2011}).

\bibitem[{\citenamefont{Deutsch and Hart}(1986)}]{PhysRevA.34.5168}
\bibinfo{author}{\bibfnamefont{M.}~\bibnamefont{Deutsch}} \bibnamefont{and}
  \bibinfo{author}{\bibfnamefont{M.}~\bibnamefont{Hart}},
  \bibinfo{journal}{Phys. Rev. A} \textbf{\bibinfo{volume}{34}},
  \bibinfo{pages}{5168} (\bibinfo{year}{1986}).

\bibitem[{\citenamefont{Kav{\v{c}}i{\v{c}}
  et~al.}(2009)\citenamefont{Kav{\v{c}}i{\v{c}}, {\v{Z}}itnik, Bu{\v{c}}ar,
  Miheli{\v{c}}, {\v{S}}tuhec, Szlachetko, Cao, Mori, and
  Glatzel}}]{Kavcic2009}
\bibinfo{author}{\bibfnamefont{M.}~\bibnamefont{Kav{\v{c}}i{\v{c}}}},
  \bibinfo{author}{\bibfnamefont{M.}~\bibnamefont{{\v{Z}}itnik}},
  \bibinfo{author}{\bibfnamefont{K.}~\bibnamefont{Bu{\v{c}}ar}},
  \bibinfo{author}{\bibfnamefont{A.}~\bibnamefont{Miheli{\v{c}}}},
  \bibinfo{author}{\bibfnamefont{M.}~\bibnamefont{{\v{S}}tuhec}},
  \bibinfo{author}{\bibfnamefont{J.}~\bibnamefont{Szlachetko}},
  \bibinfo{author}{\bibfnamefont{W.}~\bibnamefont{Cao}},
  \bibinfo{author}{\bibfnamefont{R.~A.} \bibnamefont{Mori}}, \bibnamefont{and}
  \bibinfo{author}{\bibfnamefont{P.}~\bibnamefont{Glatzel}},
  \bibinfo{journal}{Phys. Rev. Lett.} \textbf{\bibinfo{volume}{102}},
  \bibinfo{pages}{143001} (\bibinfo{year}{2009}).

\bibitem[{\citenamefont{Elton et~al.}(2000)\citenamefont{Elton, Cobble, Griem,
  Montgomery, Mancini, Jacobs, and Behar}}]{Elton2000}
\bibinfo{author}{\bibfnamefont{R.}~\bibnamefont{Elton}},
  \bibinfo{author}{\bibfnamefont{J.}~\bibnamefont{Cobble}},
  \bibinfo{author}{\bibfnamefont{H.}~\bibnamefont{Griem}},
  \bibinfo{author}{\bibfnamefont{D.}~\bibnamefont{Montgomery}},
  \bibinfo{author}{\bibfnamefont{R.}~\bibnamefont{Mancini}},
  \bibinfo{author}{\bibfnamefont{V.}~\bibnamefont{Jacobs}}, \bibnamefont{and}
  \bibinfo{author}{\bibfnamefont{E.}~\bibnamefont{Behar}}, \bibinfo{journal}{J.
  Quant. Spectrosc. Radiat. Transfer} \textbf{\bibinfo{volume}{65}},
  \bibinfo{pages}{185 } (\bibinfo{year}{2000}).

\bibitem[{\citenamefont{Lunney}(1983)}]{Lunney1983}
\bibinfo{author}{\bibfnamefont{J.~G.} \bibnamefont{Lunney}},
  \bibinfo{journal}{J. Phys. B: At. Mol. Phys.} \textbf{\bibinfo{volume}{16}},
  \bibinfo{pages}{L631} (\bibinfo{year}{1983}).

\bibitem[{\citenamefont{Bitter et~al.}(1984)\citenamefont{Bitter, von Goeler,
  Cohen, Hill, Sesnic, Tenney, Timberlake, Safronova, Vainshtein, Dubau
  et~al.}}]{PhysRevA.29.661}
\bibinfo{author}{\bibfnamefont{M.}~\bibnamefont{Bitter}},
  \bibinfo{author}{\bibfnamefont{S.}~\bibnamefont{von Goeler}},
  \bibinfo{author}{\bibfnamefont{S.}~\bibnamefont{Cohen}},
  \bibinfo{author}{\bibfnamefont{K.~W.} \bibnamefont{Hill}},
  \bibinfo{author}{\bibfnamefont{S.}~\bibnamefont{Sesnic}},
  \bibinfo{author}{\bibfnamefont{F.}~\bibnamefont{Tenney}},
  \bibinfo{author}{\bibfnamefont{J.}~\bibnamefont{Timberlake}},
  \bibinfo{author}{\bibfnamefont{U.~I.} \bibnamefont{Safronova}},
  \bibinfo{author}{\bibfnamefont{L.~A.} \bibnamefont{Vainshtein}},
  \bibinfo{author}{\bibfnamefont{J.}~\bibnamefont{Dubau}},
  \bibnamefont{et~al.}, \bibinfo{journal}{Phys. Rev. A}
  \textbf{\bibinfo{volume}{29}}, \bibinfo{pages}{661} (\bibinfo{year}{1984}).

\bibitem[{\citenamefont{Andriamonje et~al.}(1991)\citenamefont{Andriamonje,
  Andra, and Simionovici}}]{Andriamonje1991}
\bibinfo{author}{\bibfnamefont{S.}~\bibnamefont{Andriamonje}},
  \bibinfo{author}{\bibfnamefont{H.~J.} \bibnamefont{Andra}}, \bibnamefont{and}
  \bibinfo{author}{\bibfnamefont{A.}~\bibnamefont{Simionovici}},
  \bibinfo{journal}{Eur. Phys. J. D} \textbf{\bibinfo{volume}{21}},
  \bibinfo{pages}{349} (\bibinfo{year}{1991}).

\bibitem[{\citenamefont{Porquet et~al.}(2010)\citenamefont{Porquet, Dubau, and
  Grosso}}]{Porquet2010}
\bibinfo{author}{\bibfnamefont{D.}~\bibnamefont{Porquet}},
  \bibinfo{author}{\bibfnamefont{J.}~\bibnamefont{Dubau}}, \bibnamefont{and}
  \bibinfo{author}{\bibfnamefont{N.}~\bibnamefont{Grosso}},
  \bibinfo{journal}{Space Sci. Rev.} \textbf{\bibinfo{volume}{157}},
  \bibinfo{pages}{103} (\bibinfo{year}{2010}).

\bibitem[{\citenamefont{Decaux et~al.}(1997)\citenamefont{Decaux, Beiersdorfer,
  Kahn, and Jacobs}}]{Decaux1997}
\bibinfo{author}{\bibfnamefont{V.}~\bibnamefont{Decaux}},
  \bibinfo{author}{\bibfnamefont{P.}~\bibnamefont{Beiersdorfer}},
  \bibinfo{author}{\bibfnamefont{S.~M.} \bibnamefont{Kahn}}, \bibnamefont{and}
  \bibinfo{author}{\bibfnamefont{V.~L.} \bibnamefont{Jacobs}},
  \bibinfo{journal}{The Astrophysical Journal} \textbf{\bibinfo{volume}{482}},
  \bibinfo{pages}{1076} (\bibinfo{year}{1997}).

\bibitem[{\citenamefont{Costa et~al.}(2006{\natexlab{a}})\citenamefont{Costa,
  Martins, Santos, Indelicato, and Parente}}]{Costa2006a}
\bibinfo{author}{\bibfnamefont{A.}~\bibnamefont{Costa}},
  \bibinfo{author}{\bibfnamefont{M.}~\bibnamefont{Martins}},
  \bibinfo{author}{\bibfnamefont{J.}~\bibnamefont{Santos}},
  \bibinfo{author}{\bibfnamefont{P.}~\bibnamefont{Indelicato}},
  \bibnamefont{and} \bibinfo{author}{\bibfnamefont{F.}~\bibnamefont{Parente}},
  \bibinfo{journal}{J. Phys. B: At., Mol. Opt. Phys.}
  \textbf{\bibinfo{volume}{39}}, \bibinfo{pages}{2355}
  (\bibinfo{year}{2006}{\natexlab{a}}).

\bibitem[{\citenamefont{Heisenberg}(1925)}]{Heisenberg1925}
\bibinfo{author}{\bibfnamefont{W.}~\bibnamefont{Heisenberg}},
  \bibinfo{journal}{Zeitschrift f{\"u}r Physik} \textbf{\bibinfo{volume}{32}},
  \bibinfo{pages}{841} (\bibinfo{year}{1925}).

\bibitem[{\citenamefont{W\"{o}lfli et~al.}(1975)\citenamefont{W\"{o}lfli,
  Stoller, Bonani, Suter, and St\"ockli}}]{PhysRevLett.35.656}
\bibinfo{author}{\bibfnamefont{W.}~\bibnamefont{W\"{o}lfli}},
  \bibinfo{author}{\bibfnamefont{C.}~\bibnamefont{Stoller}},
  \bibinfo{author}{\bibfnamefont{G.}~\bibnamefont{Bonani}},
  \bibinfo{author}{\bibfnamefont{M.}~\bibnamefont{Suter}}, \bibnamefont{and}
  \bibinfo{author}{\bibfnamefont{M.}~\bibnamefont{St\"ockli}},
  \bibinfo{journal}{Phys. Rev. Lett.} \textbf{\bibinfo{volume}{35}},
  \bibinfo{pages}{656} (\bibinfo{year}{1975}).

\bibitem[{\citenamefont{Ding et~al.}(2020)\citenamefont{Ding, Wu, Cao, Zhang,
  Zhang, Xue, Yu, and Dong}}]{Ding2020}
\bibinfo{author}{\bibfnamefont{X.}~\bibnamefont{Ding}},
  \bibinfo{author}{\bibfnamefont{C.}~\bibnamefont{Wu}},
  \bibinfo{author}{\bibfnamefont{M.}~\bibnamefont{Cao}},
  \bibinfo{author}{\bibfnamefont{D.}~\bibnamefont{Zhang}},
  \bibinfo{author}{\bibfnamefont{M.}~\bibnamefont{Zhang}},
  \bibinfo{author}{\bibfnamefont{Y.}~\bibnamefont{Xue}},
  \bibinfo{author}{\bibfnamefont{D.}~\bibnamefont{Yu}}, \bibnamefont{and}
  \bibinfo{author}{\bibfnamefont{C.}~\bibnamefont{Dong}},
  \bibinfo{journal}{Chin. Phys. B} \textbf{\bibinfo{volume}{29}},
  \bibinfo{eid}{33101} (\bibinfo{year}{2020}).

\bibitem[{\citenamefont{Kozio{\l} and
  Rzadkiewicz}(2017)}]{koziol2017theoretical}
\bibinfo{author}{\bibfnamefont{K.}~\bibnamefont{Kozio{\l}}} \bibnamefont{and}
  \bibinfo{author}{\bibfnamefont{J.}~\bibnamefont{Rzadkiewicz}},
  \bibinfo{journal}{Phys. Rev. A} \textbf{\bibinfo{volume}{96}},
  \bibinfo{pages}{031402} (\bibinfo{year}{2017}).

\bibitem[{\citenamefont{Costa et~al.}(2006{\natexlab{b}})\citenamefont{Costa,
  Martins, Santos, Indelicato, and Parente}}]{Costa2006}
\bibinfo{author}{\bibfnamefont{A.}~\bibnamefont{Costa}},
  \bibinfo{author}{\bibfnamefont{M.}~\bibnamefont{Martins}},
  \bibinfo{author}{\bibfnamefont{J.}~\bibnamefont{Santos}},
  \bibinfo{author}{\bibfnamefont{P.}~\bibnamefont{Indelicato}},
  \bibnamefont{and} \bibinfo{author}{\bibfnamefont{F.}~\bibnamefont{Parente}},
  \bibinfo{journal}{J. Phys. B: At., Mol. Opt. Phys.}
  \textbf{\bibinfo{volume}{39}}, \bibinfo{pages}{2355}
  (\bibinfo{year}{2006}{\natexlab{b}}).

\bibitem[{\citenamefont{Safronova and
  Senachenko}(1981)}]{safronova1981influence}
\bibinfo{author}{\bibfnamefont{U.}~\bibnamefont{Safronova}} \bibnamefont{and}
  \bibinfo{author}{\bibfnamefont{V.}~\bibnamefont{Senachenko}},
  \bibinfo{journal}{J. Phys. B: At. Mol. Phys.} \textbf{\bibinfo{volume}{14}},
  \bibinfo{pages}{603} (\bibinfo{year}{1981}).

\bibitem[{\citenamefont{Martins et~al.}(2004)\citenamefont{Martins, Costa,
  Santos, Parente, and Indelicato}}]{Martins2004}
\bibinfo{author}{\bibfnamefont{M.}~\bibnamefont{Martins}},
  \bibinfo{author}{\bibfnamefont{A.}~\bibnamefont{Costa}},
  \bibinfo{author}{\bibfnamefont{J.}~\bibnamefont{Santos}},
  \bibinfo{author}{\bibfnamefont{F.}~\bibnamefont{Parente}}, \bibnamefont{and}
  \bibinfo{author}{\bibfnamefont{P.}~\bibnamefont{Indelicato}},
  \bibinfo{journal}{J. Phys. B: At., Mol. Opt. Phys.}
  \textbf{\bibinfo{volume}{37}}, \bibinfo{pages}{3785} (\bibinfo{year}{2004}).

\bibitem[{\citenamefont{Santos et~al.}(2003)\citenamefont{Santos, Martins,
  Costa, Parente, and Indelicato}}]{santos2003two}
\bibinfo{author}{\bibfnamefont{J.}~\bibnamefont{Santos}},
  \bibinfo{author}{\bibfnamefont{M.}~\bibnamefont{Martins}},
  \bibinfo{author}{\bibfnamefont{A.}~\bibnamefont{Costa}},
  \bibinfo{author}{\bibfnamefont{F.}~\bibnamefont{Parente}}, \bibnamefont{and}
  \bibinfo{author}{\bibfnamefont{P.}~\bibnamefont{Indelicato}},
  \bibinfo{journal}{Nucl. Instrum. Methods Phys. Res., Sect. B}
  \textbf{\bibinfo{volume}{205}}, \bibinfo{pages}{102} (\bibinfo{year}{2003}).

\bibitem[{\citenamefont{Saha et~al.}(2009)\citenamefont{Saha, Mukherjee,
  Fritzsche, and Mukherjee}}]{saha2009effect}
\bibinfo{author}{\bibfnamefont{J.}~\bibnamefont{Saha}},
  \bibinfo{author}{\bibfnamefont{T.}~\bibnamefont{Mukherjee}},
  \bibinfo{author}{\bibfnamefont{S.}~\bibnamefont{Fritzsche}},
  \bibnamefont{and}
  \bibinfo{author}{\bibfnamefont{P.}~\bibnamefont{Mukherjee}},
  \bibinfo{journal}{Phys. Lett. A} \textbf{\bibinfo{volume}{373}},
  \bibinfo{pages}{252} (\bibinfo{year}{2009}).

\bibitem[{\citenamefont{Natarajan}(2008)}]{Natarajan2008}
\bibinfo{author}{\bibfnamefont{L.}~\bibnamefont{Natarajan}},
  \bibinfo{journal}{Phys. Rev. A} \textbf{\bibinfo{volume}{78}},
  \bibinfo{pages}{052505} (\bibinfo{year}{2008}).

\bibitem[{\citenamefont{{\AA}berg et~al.}(1976)\citenamefont{{\AA}berg,
  Jamison, and Richard}}]{aaberg1976origin}
\bibinfo{author}{\bibfnamefont{T.}~\bibnamefont{{\AA}berg}},
  \bibinfo{author}{\bibfnamefont{K.}~\bibnamefont{Jamison}}, \bibnamefont{and}
  \bibinfo{author}{\bibfnamefont{P.}~\bibnamefont{Richard}},
  \bibinfo{journal}{Phys. Rev. Lett.} \textbf{\bibinfo{volume}{37}},
  \bibinfo{pages}{63} (\bibinfo{year}{1976}).

\bibitem[{\citenamefont{Kadrekar and
  Natarajan}(2011{\natexlab{a}})}]{kadrekar2011radiative}
\bibinfo{author}{\bibfnamefont{R.}~\bibnamefont{Kadrekar}} \bibnamefont{and}
  \bibinfo{author}{\bibfnamefont{L.}~\bibnamefont{Natarajan}},
  \bibinfo{journal}{Phys. Rev. A} \textbf{\bibinfo{volume}{84}},
  \bibinfo{pages}{062506} (\bibinfo{year}{2011}{\natexlab{a}}).

\bibitem[{\citenamefont{Shao et~al.}(2017)\citenamefont{Shao, Yu, Cai, Chen,
  Ma, Evslin, Xue, Wang, Kozhedub, Lu et~al.}}]{Shao2017}
\bibinfo{author}{\bibfnamefont{C.}~\bibnamefont{Shao}},
  \bibinfo{author}{\bibfnamefont{D.}~\bibnamefont{Yu}},
  \bibinfo{author}{\bibfnamefont{X.}~\bibnamefont{Cai}},
  \bibinfo{author}{\bibfnamefont{X.}~\bibnamefont{Chen}},
  \bibinfo{author}{\bibfnamefont{K.}~\bibnamefont{Ma}},
  \bibinfo{author}{\bibfnamefont{J.}~\bibnamefont{Evslin}},
  \bibinfo{author}{\bibfnamefont{Y.}~\bibnamefont{Xue}},
  \bibinfo{author}{\bibfnamefont{W.}~\bibnamefont{Wang}},
  \bibinfo{author}{\bibfnamefont{Y.~S.} \bibnamefont{Kozhedub}},
  \bibinfo{author}{\bibfnamefont{R.}~\bibnamefont{Lu}}, \bibnamefont{et~al.},
  \bibinfo{journal}{Phys. Rev. A} \textbf{\bibinfo{volume}{96}},
  \bibinfo{pages}{012708} (\bibinfo{year}{2017}).

\bibitem[{\citenamefont{Jabłoński et~al.}(2020)\citenamefont{Jabłoński,
  Banaś, Jagodziński, Kubala-Kukuś, Sobota, Stabrawa, Szary, and
  Pajek}}]{Jablonski2020a}
\bibinfo{author}{\bibfnamefont{{\L}.}~\bibnamefont{Jabłoński}},
  \bibinfo{author}{\bibfnamefont{D.}~\bibnamefont{Banaś}},
  \bibinfo{author}{\bibfnamefont{P.}~\bibnamefont{Jagodziński}},
  \bibinfo{author}{\bibfnamefont{A.}~\bibnamefont{Kubala-Kukuś}},
  \bibinfo{author}{\bibfnamefont{D.}~\bibnamefont{Sobota}},
  \bibinfo{author}{\bibfnamefont{I.}~\bibnamefont{Stabrawa}},
  \bibinfo{author}{\bibfnamefont{K.}~\bibnamefont{Szary}}, \bibnamefont{and}
  \bibinfo{author}{\bibfnamefont{M.}~\bibnamefont{Pajek}}, \bibinfo{journal}{J.
  Phys. Conf. Ser.} \textbf{\bibinfo{volume}{1412}}, \bibinfo{pages}{202002}
  (\bibinfo{year}{2020}).

\bibitem[{\citenamefont{Ding et~al.}(2011)\citenamefont{Ding, Koike, Murakami,
  Kato, Sakaue, Dong, Nakamura, Komatsu, and Sakoda}}]{Ding2011}
\bibinfo{author}{\bibfnamefont{X.~B.} \bibnamefont{Ding}},
  \bibinfo{author}{\bibfnamefont{F.}~\bibnamefont{Koike}},
  \bibinfo{author}{\bibfnamefont{I.}~\bibnamefont{Murakami}},
  \bibinfo{author}{\bibfnamefont{D.}~\bibnamefont{Kato}},
  \bibinfo{author}{\bibfnamefont{H.~A.} \bibnamefont{Sakaue}},
  \bibinfo{author}{\bibfnamefont{C.~Z.} \bibnamefont{Dong}},
  \bibinfo{author}{\bibfnamefont{N.}~\bibnamefont{Nakamura}},
  \bibinfo{author}{\bibfnamefont{A.}~\bibnamefont{Komatsu}}, \bibnamefont{and}
  \bibinfo{author}{\bibfnamefont{J.}~\bibnamefont{Sakoda}},
  \bibinfo{journal}{J. Phys. B: At. Mol. Opt. Phys.}
  \textbf{\bibinfo{volume}{44}}, \bibinfo{pages}{145004}
  (\bibinfo{year}{2011}).

\bibitem[{\citenamefont{Ding et~al.}(2017)\citenamefont{Ding, Sun, Liu, Koike,
  Murakami, Kato, Sakaue, Nakamura, and Dong}}]{Ding2017a}
\bibinfo{author}{\bibfnamefont{X.~B.} \bibnamefont{Ding}},
  \bibinfo{author}{\bibfnamefont{R.}~\bibnamefont{Sun}},
  \bibinfo{author}{\bibfnamefont{J.~X.} \bibnamefont{Liu}},
  \bibinfo{author}{\bibfnamefont{F.}~\bibnamefont{Koike}},
  \bibinfo{author}{\bibfnamefont{I.}~\bibnamefont{Murakami}},
  \bibinfo{author}{\bibfnamefont{D.}~\bibnamefont{Kato}},
  \bibinfo{author}{\bibfnamefont{H.}~\bibnamefont{Sakaue}},
  \bibinfo{author}{\bibfnamefont{N.}~\bibnamefont{Nakamura}}, \bibnamefont{and}
  \bibinfo{author}{\bibfnamefont{C.~Z.} \bibnamefont{Dong}},
  \bibinfo{journal}{J. Phys. B: At., Mol. Opt. Phys.}
  \textbf{\bibinfo{volume}{50}}, \bibinfo{pages}{045004}
  (\bibinfo{year}{2017}).

\bibitem[{\citenamefont{Aggarwal and Keenan}(2016)}]{Aggarwal2016187}
\bibinfo{author}{\bibfnamefont{K.~M.} \bibnamefont{Aggarwal}} \bibnamefont{and}
  \bibinfo{author}{\bibfnamefont{F.~P.} \bibnamefont{Keenan}},
  \bibinfo{journal}{At. Data Nucl. Data Tables}
  \textbf{\bibinfo{volume}{111-112}}, \bibinfo{pages}{187 }
  (\bibinfo{year}{2016}).

\bibitem[{\citenamefont{Kadrekar and
  Natarajan}(2011{\natexlab{b}})}]{PhysRevA.84.062506}
\bibinfo{author}{\bibfnamefont{R.}~\bibnamefont{Kadrekar}} \bibnamefont{and}
  \bibinfo{author}{\bibfnamefont{L.}~\bibnamefont{Natarajan}},
  \bibinfo{journal}{Phys. Rev. A} \textbf{\bibinfo{volume}{84}},
  \bibinfo{pages}{062506} (\bibinfo{year}{2011}{\natexlab{b}}).

\bibitem[{\citenamefont{Natarajan}(2013)}]{natarajan2013two}
\bibinfo{author}{\bibfnamefont{L.}~\bibnamefont{Natarajan}},
  \bibinfo{journal}{Phys. Rev. A} \textbf{\bibinfo{volume}{88}},
  \bibinfo{pages}{052522} (\bibinfo{year}{2013}).

\bibitem[{\citenamefont{Grant}(2007)}]{Grant2007}
\bibinfo{author}{\bibfnamefont{I.~P.} \bibnamefont{Grant}},
  \emph{\bibinfo{title}{Relativistic Quantum Theory of Atoms and Molecules,
  Theory and Computation}} (\bibinfo{publisher}{Springer, New York},
  \bibinfo{year}{2007}).

\bibitem[{\citenamefont{Grant et~al.}(1980)\citenamefont{Grant, McKenzie,
  Norrington, Mayers, and Pyper}}]{Grant1980}
\bibinfo{author}{\bibfnamefont{I.~P.} \bibnamefont{Grant}},
  \bibinfo{author}{\bibfnamefont{B.~J.} \bibnamefont{McKenzie}},
  \bibinfo{author}{\bibfnamefont{P.~H.} \bibnamefont{Norrington}},
  \bibinfo{author}{\bibfnamefont{D.~F.} \bibnamefont{Mayers}},
  \bibnamefont{and} \bibinfo{author}{\bibfnamefont{N.~C.} \bibnamefont{Pyper}},
  \bibinfo{journal}{Comput. Phys. Commun.} \textbf{\bibinfo{volume}{21}},
  \bibinfo{pages}{207} (\bibinfo{year}{1980}).

\bibitem[{\citenamefont{McKenzie et~al.}(1980)\citenamefont{McKenzie, Grant,
  and Norrington}}]{McKenzie1980}
\bibinfo{author}{\bibfnamefont{B.~J.} \bibnamefont{McKenzie}},
  \bibinfo{author}{\bibfnamefont{I.~P.} \bibnamefont{Grant}}, \bibnamefont{and}
  \bibinfo{author}{\bibfnamefont{P.~H.} \bibnamefont{Norrington}},
  \bibinfo{journal}{Comput. Phys. Commun.} \textbf{\bibinfo{volume}{21}},
  \bibinfo{pages}{233} (\bibinfo{year}{1980}).

\bibitem[{\citenamefont{Dyall et~al.}(1989)\citenamefont{Dyall, Grant, Johnson,
  Parpia, and Plummer}}]{K.1989}
\bibinfo{author}{\bibfnamefont{K.~G.} \bibnamefont{Dyall}},
  \bibinfo{author}{\bibfnamefont{I.~P.} \bibnamefont{Grant}},
  \bibinfo{author}{\bibfnamefont{C.~T.} \bibnamefont{Johnson}},
  \bibinfo{author}{\bibfnamefont{F.~A.} \bibnamefont{Parpia}},
  \bibnamefont{and} \bibinfo{author}{\bibfnamefont{E.~P.}
  \bibnamefont{Plummer}}, \bibinfo{journal}{Comput. Phys. Commun.}
  \textbf{\bibinfo{volume}{55}}, \bibinfo{pages}{425} (\bibinfo{year}{1989}).

\bibitem[{\citenamefont{Parpia et~al.}(1996)\citenamefont{Parpia, Fischer, and
  Grant}}]{GRASP92}
\bibinfo{author}{\bibfnamefont{F.~A.} \bibnamefont{Parpia}},
  \bibinfo{author}{\bibfnamefont{C.~F.} \bibnamefont{Fischer}},
  \bibnamefont{and} \bibinfo{author}{\bibfnamefont{I.~P.} \bibnamefont{Grant}},
  \bibinfo{journal}{Comput. Phys. Commun.} \textbf{\bibinfo{volume}{94}},
  \bibinfo{pages}{249} (\bibinfo{year}{1996}).

\bibitem[{\citenamefont{J\"{o}nsson et~al.}(2007)\citenamefont{J\"{o}nsson, He,
  Fischer, and Grant}}]{Jonsson2007}
\bibinfo{author}{\bibfnamefont{P.}~\bibnamefont{J\"{o}nsson}},
  \bibinfo{author}{\bibfnamefont{X.}~\bibnamefont{He}},
  \bibinfo{author}{\bibfnamefont{C.~F.} \bibnamefont{Fischer}},
  \bibnamefont{and} \bibinfo{author}{\bibfnamefont{I.}~\bibnamefont{Grant}},
  \bibinfo{journal}{Comput. Phys. Commun.} \textbf{\bibinfo{volume}{177}},
  \bibinfo{pages}{597} (\bibinfo{year}{2007}).

\bibitem[{\citenamefont{Jönsson et~al.}(2013)\citenamefont{Jönsson, Gaigalas,
  Biero{\'{n}}, Fischer, and Grant}}]{Joensson2013a}
\bibinfo{author}{\bibfnamefont{P.}~\bibnamefont{Jönsson}},
  \bibinfo{author}{\bibfnamefont{G.}~\bibnamefont{Gaigalas}},
  \bibinfo{author}{\bibfnamefont{J.}~\bibnamefont{Biero{\'{n}}}},
  \bibinfo{author}{\bibfnamefont{C.~F.} \bibnamefont{Fischer}},
  \bibnamefont{and} \bibinfo{author}{\bibfnamefont{I.}~\bibnamefont{Grant}},
  \bibinfo{journal}{Comput. Phys. Commun.} \textbf{\bibinfo{volume}{184}},
  \bibinfo{pages}{2197} (\bibinfo{year}{2013}).

\bibitem[{\citenamefont{Fischer et~al.}(2019)\citenamefont{Fischer, Gaigalas,
  Jönsson, and Biero{\'{n}}}}]{Fischer2019}
\bibinfo{author}{\bibfnamefont{C.~F.} \bibnamefont{Fischer}},
  \bibinfo{author}{\bibfnamefont{G.}~\bibnamefont{Gaigalas}},
  \bibinfo{author}{\bibfnamefont{P.}~\bibnamefont{Jönsson}}, \bibnamefont{and}
  \bibinfo{author}{\bibfnamefont{J.}~\bibnamefont{Biero{\'{n}}}},
  \bibinfo{journal}{Comput. Phys. Commun.} \textbf{\bibinfo{volume}{237}},
  \bibinfo{pages}{184} (\bibinfo{year}{2019}).

\bibitem[{\citenamefont{Cheng et~al.}(1979)\citenamefont{Cheng, Kim, and
  Desclaux}}]{Cheng1979}
\bibinfo{author}{\bibfnamefont{K.~T.} \bibnamefont{Cheng}},
  \bibinfo{author}{\bibfnamefont{Y.}~\bibnamefont{Kim}}, \bibnamefont{and}
  \bibinfo{author}{\bibfnamefont{J.~P.} \bibnamefont{Desclaux}},
  \bibinfo{journal}{At. Data Nucl. Data Tables} \textbf{\bibinfo{volume}{24}},
  \bibinfo{pages}{111} (\bibinfo{year}{1979}).

\bibitem[{\citenamefont{Kramida et~al.}(2019)\citenamefont{Kramida,
  {Yu.~Ralchenko}, Reader, and {and NIST ASD Team}}}]{NIST_ASD}
\bibinfo{author}{\bibfnamefont{A.}~\bibnamefont{Kramida}},
  \bibinfo{author}{\bibnamefont{{Yu.~Ralchenko}}},
  \bibinfo{author}{\bibfnamefont{J.}~\bibnamefont{Reader}}, \bibnamefont{and}
  \bibinfo{author}{\bibnamefont{{and NIST ASD Team}}},
  \bibinfo{howpublished}{{NIST Atomic Spectra Database (ver. 5.7.1), [Online].
  Available: {\tt{https://physics.nist.gov/asd}} [2020, July 25]. National
  Institute of Standards and Technology, Gaithersburg, MD.}}
  (\bibinfo{year}{2019}).

\bibitem[{\citenamefont{Aggarwal et~al.}(2010)\citenamefont{Aggarwal, Keenan,
  and Lawson}}]{aggarwal2010energy}
\bibinfo{author}{\bibfnamefont{K.}~\bibnamefont{Aggarwal}},
  \bibinfo{author}{\bibfnamefont{F.}~\bibnamefont{Keenan}}, \bibnamefont{and}
  \bibinfo{author}{\bibfnamefont{K.}~\bibnamefont{Lawson}},
  \bibinfo{journal}{At. Data Nucl. Data Tables} \textbf{\bibinfo{volume}{96}},
  \bibinfo{pages}{123} (\bibinfo{year}{2010}).

\bibitem[{\citenamefont{Li et~al.}(2010)\citenamefont{Li, Jönsson, Dong, and
  Gaigalas}}]{0953-4075-43-3-035005}
\bibinfo{author}{\bibfnamefont{J.}~\bibnamefont{Li}},
  \bibinfo{author}{\bibfnamefont{P.}~\bibnamefont{Jönsson}},
  \bibinfo{author}{\bibfnamefont{C.}~\bibnamefont{Dong}}, \bibnamefont{and}
  \bibinfo{author}{\bibfnamefont{G.}~\bibnamefont{Gaigalas}},
  \bibinfo{journal}{J. Phys. B: At., Mol. Opt. Phys.}
  \textbf{\bibinfo{volume}{43}}, \bibinfo{pages}{035005}
  (\bibinfo{year}{2010}).

\bibitem[{\citenamefont{Plante et~al.}(1994)\citenamefont{Plante, Johnson, and
  Sapirstein}}]{plante1994relativistic}
\bibinfo{author}{\bibfnamefont{D.}~\bibnamefont{Plante}},
  \bibinfo{author}{\bibfnamefont{W.}~\bibnamefont{Johnson}}, \bibnamefont{and}
  \bibinfo{author}{\bibfnamefont{J.}~\bibnamefont{Sapirstein}},
  \bibinfo{journal}{Phys. Rev. A} \textbf{\bibinfo{volume}{49}},
  \bibinfo{pages}{3519} (\bibinfo{year}{1994}).

\bibitem[{\citenamefont{Cowan}(1981)}]{Cowan1981}
\bibinfo{author}{\bibfnamefont{R.}~\bibnamefont{Cowan}}, \bibinfo{journal}{Los
  Alamos Series in Basic and Applied Sciences, Berkeley: University of
  California Press, 1981}  (\bibinfo{year}{1981}).

\bibitem[{\citenamefont{Elsayed et~al.}(2015)\citenamefont{Elsayed, Khered, and
  Attia}}]{Elsayed2015}
\bibinfo{author}{\bibfnamefont{F.}~\bibnamefont{Elsayed}},
  \bibinfo{author}{\bibfnamefont{M.}~\bibnamefont{Khered}}, \bibnamefont{and}
  \bibinfo{author}{\bibfnamefont{S.~M.} \bibnamefont{Attia}},
  \bibinfo{journal}{Eur. Phys. J. P} \textbf{\bibinfo{volume}{130}},
  \bibinfo{pages}{104} (\bibinfo{year}{2015}).

\bibitem[{\citenamefont{Uhlig et~al.}(2015)\citenamefont{Uhlig, Doriese,
  Fowler, Swetz, Jaye, Fischer, Reintsema, Bennett, Vale, Mandal
  et~al.}}]{uhlig2015high}
\bibinfo{author}{\bibfnamefont{J.}~\bibnamefont{Uhlig}},
  \bibinfo{author}{\bibfnamefont{W.}~\bibnamefont{Doriese}},
  \bibinfo{author}{\bibfnamefont{J.}~\bibnamefont{Fowler}},
  \bibinfo{author}{\bibfnamefont{D.}~\bibnamefont{Swetz}},
  \bibinfo{author}{\bibfnamefont{C.}~\bibnamefont{Jaye}},
  \bibinfo{author}{\bibfnamefont{D.}~\bibnamefont{Fischer}},
  \bibinfo{author}{\bibfnamefont{C.}~\bibnamefont{Reintsema}},
  \bibinfo{author}{\bibfnamefont{D.}~\bibnamefont{Bennett}},
  \bibinfo{author}{\bibfnamefont{L.}~\bibnamefont{Vale}},
  \bibinfo{author}{\bibfnamefont{U.}~\bibnamefont{Mandal}},
  \bibnamefont{et~al.}, \bibinfo{journal}{J. Synchrotron Rad.}
  \textbf{\bibinfo{volume}{22}}, \bibinfo{pages}{766} (\bibinfo{year}{2015}).

\bibitem[{\citenamefont{Briand et~al.}(1990)\citenamefont{Briand, De~Billy,
  Charles, Essabaa, Briand, Geller, Desclaux, Bliman, and
  Ristori}}]{briand1990production}
\bibinfo{author}{\bibfnamefont{J.}~\bibnamefont{Briand}},
  \bibinfo{author}{\bibfnamefont{L.}~\bibnamefont{De~Billy}},
  \bibinfo{author}{\bibfnamefont{P.}~\bibnamefont{Charles}},
  \bibinfo{author}{\bibfnamefont{S.}~\bibnamefont{Essabaa}},
  \bibinfo{author}{\bibfnamefont{P.}~\bibnamefont{Briand}},
  \bibinfo{author}{\bibfnamefont{R.}~\bibnamefont{Geller}},
  \bibinfo{author}{\bibfnamefont{J.}~\bibnamefont{Desclaux}},
  \bibinfo{author}{\bibfnamefont{S.}~\bibnamefont{Bliman}}, \bibnamefont{and}
  \bibinfo{author}{\bibfnamefont{C.}~\bibnamefont{Ristori}},
  \bibinfo{journal}{Phys. Rev. Lett.} \textbf{\bibinfo{volume}{65}},
  \bibinfo{pages}{159} (\bibinfo{year}{1990}).

\bibitem[{\citenamefont{Winter and Aumayr}(1999)}]{winter1999hollow}
\bibinfo{author}{\bibfnamefont{H.}~\bibnamefont{Winter}} \bibnamefont{and}
  \bibinfo{author}{\bibfnamefont{F.}~\bibnamefont{Aumayr}},
  \bibinfo{journal}{J. Phys. B: At., Mol. Opt. Phys.}
  \textbf{\bibinfo{volume}{32}}, \bibinfo{pages}{R39} (\bibinfo{year}{1999}).

\bibitem[{\citenamefont{Briand et~al.}(1989)\citenamefont{Briand, Indelicato,
  Simionovici, Vicente, Liesen, and Dietrich}}]{Briand_1989}
\bibinfo{author}{\bibfnamefont{J.~P.} \bibnamefont{Briand}},
  \bibinfo{author}{\bibfnamefont{P.}~\bibnamefont{Indelicato}},
  \bibinfo{author}{\bibfnamefont{A.}~\bibnamefont{Simionovici}},
  \bibinfo{author}{\bibfnamefont{V.~S.} \bibnamefont{Vicente}},
  \bibinfo{author}{\bibfnamefont{D.}~\bibnamefont{Liesen}}, \bibnamefont{and}
  \bibinfo{author}{\bibfnamefont{D.}~\bibnamefont{Dietrich}},
  \bibinfo{journal}{Europhysics Letters ({EPL})} \textbf{\bibinfo{volume}{9}},
  \bibinfo{pages}{225} (\bibinfo{year}{1989}).

\bibitem[{\citenamefont{Lamour et~al.}(2015)\citenamefont{Lamour, Fainstein,
  Galassi, Prigent, Ramirez, Rivarola, Rozet, Trassinelli, and
  Vernhet}}]{LamourExtension}
\bibinfo{author}{\bibfnamefont{E.}~\bibnamefont{Lamour}},
  \bibinfo{author}{\bibfnamefont{P.~D.} \bibnamefont{Fainstein}},
  \bibinfo{author}{\bibfnamefont{M.}~\bibnamefont{Galassi}},
  \bibinfo{author}{\bibfnamefont{C.}~\bibnamefont{Prigent}},
  \bibinfo{author}{\bibfnamefont{C.~A.} \bibnamefont{Ramirez}},
  \bibinfo{author}{\bibfnamefont{R.~D.} \bibnamefont{Rivarola}},
  \bibinfo{author}{\bibfnamefont{J.-P.} \bibnamefont{Rozet}},
  \bibinfo{author}{\bibfnamefont{M.}~\bibnamefont{Trassinelli}},
  \bibnamefont{and} \bibinfo{author}{\bibfnamefont{D.}~\bibnamefont{Vernhet}},
  \bibinfo{journal}{Phys. Rev. A} \textbf{\bibinfo{volume}{92}},
  \bibinfo{pages}{042703} (\bibinfo{year}{2015}).

\bibitem[{\citenamefont{Vockenhuber et~al.}(2013)\citenamefont{Vockenhuber,
  Jensen, Julin, Kettunen, Laitinen, Rossi, Sajavaara, Osmani, Schinner,
  Sigmund et~al.}}]{Vockenhuber2013}
\bibinfo{author}{\bibfnamefont{C.}~\bibnamefont{Vockenhuber}},
  \bibinfo{author}{\bibfnamefont{J.}~\bibnamefont{Jensen}},
  \bibinfo{author}{\bibfnamefont{J.}~\bibnamefont{Julin}},
  \bibinfo{author}{\bibfnamefont{H.}~\bibnamefont{Kettunen}},
  \bibinfo{author}{\bibfnamefont{M.}~\bibnamefont{Laitinen}},
  \bibinfo{author}{\bibfnamefont{M.}~\bibnamefont{Rossi}},
  \bibinfo{author}{\bibfnamefont{T.}~\bibnamefont{Sajavaara}},
  \bibinfo{author}{\bibfnamefont{O.}~\bibnamefont{Osmani}},
  \bibinfo{author}{\bibfnamefont{A.}~\bibnamefont{Schinner}},
  \bibinfo{author}{\bibfnamefont{P.}~\bibnamefont{Sigmund}},
  \bibnamefont{et~al.}, \bibinfo{journal}{Eur. Phys. J. D}
  \textbf{\bibinfo{volume}{67}}, \bibinfo{pages}{145} (\bibinfo{year}{2013}).

\bibitem[{\citenamefont{Rozet et~al.}(1996)\citenamefont{Rozet, Stephan, and
  Vernhet}}]{rozet1996etacha}
\bibinfo{author}{\bibfnamefont{J.}~\bibnamefont{Rozet}},
  \bibinfo{author}{\bibfnamefont{C.}~\bibnamefont{Stephan}}, \bibnamefont{and}
  \bibinfo{author}{\bibfnamefont{D.}~\bibnamefont{Vernhet}},
  \bibinfo{journal}{Nucl. Instrum. Methods Phys. Res., Sect. B}
  \textbf{\bibinfo{volume}{107}}, \bibinfo{pages}{67} (\bibinfo{year}{1996}).

\bibitem[{\citenamefont{Tarasov and Bazin}(2016)}]{tarasov2016lise++}
\bibinfo{author}{\bibfnamefont{O.}~\bibnamefont{Tarasov}} \bibnamefont{and}
  \bibinfo{author}{\bibfnamefont{D.}~\bibnamefont{Bazin}},
  \bibinfo{journal}{Nucl. Instrum. Methods Phys. Res., Sect. B}
  \textbf{\bibinfo{volume}{376}}, \bibinfo{pages}{185} (\bibinfo{year}{2016}).

\bibitem[{\citenamefont{Tarasov et~al.}(1980)\citenamefont{Tarasov, Bazin, and
  NIM}}]{tarasov1980266}
\bibinfo{author}{\bibfnamefont{O.}~\bibnamefont{Tarasov}},
  \bibinfo{author}{\bibfnamefont{D.}~\bibnamefont{Bazin}}, \bibnamefont{and}
  \bibinfo{author}{\bibfnamefont{B.}~\bibnamefont{NIM}},
  \bibinfo{journal}{Phys. Rev. C} \textbf{\bibinfo{volume}{21}},
  \bibinfo{pages}{230} (\bibinfo{year}{1980}).

\end{thebibliography}

\end{document}